\documentstyle{article} 
\newcommand{\D}{\displaystyle}

\begin{document}					

\large

\begin{center} 
\bf 
\noindent SOLITON-COMPLEX DYNAMICS IN STRONGLY DISPERSIVE MEDIUM  

\bigskip

\bigskip

\large
\rm
Mikhail M. BOGDAN and Arnold M. KOSEVICH
\medskip 

\small
\noindent B.I.Verkin Institute for Low Temperature Physics and Engineering
of National Academy of Sciences of Ukraine,
47, Lenin Ave., 310164, Kharkov, Ukraine.

\bigskip
\large
\rm
Gerard A. MAUGIN 

\medskip 
\small
\noindent Laboratoire de Mod\'{e}lisation en M\'{e}canique, Universit\'{e} 
Pierre et Marie Curie, Boite 162, Tour 66, 4 Place Jussieu, 75252 Paris 
Cedex 05, France

\end{center}

\bigskip 

\bigskip

\normalsize\noindent \bf {Abstract.} \small 
The concept of soliton complex in a nonlinear dispersive medium is
proposed. It is shown that strongly interacting identical topological
solitons in the medium can form bound soliton complexes which move without
radiation. This phenomenon is considered to be universal and applicable to
various physical systems. The soliton complex and its "excited" states are
described analytically and numerically as solutions of nonlinear
dispersive equations with the fourth and higher order spatial or mixed
derivatives. The dispersive sine-Gordon, double and triple sine-Gordon,
and piecewise-linear models are studied in detail. Mechanisms and
conditions of the formation of soliton complexes, and peculiarities of
their stationary dynamics are investigated. A phenomenological approach to
the description of the complexes and the classification of all the
possible complex states are proposed. Some examples of physical systems,
where the phenomenon can be experimentally observed, are briefly
discussed.

\bigskip

\bigskip

\normalsize\noindent PACS number(s): 02.70.-c, 03.40.Kf, 47.20.Ky, 
47.35.+i, 62.30.+d.

\bigskip

\bigskip

\vspace*{10mm}

\begin{center}
\normalsize {\bf I. INTRODUCTION}
\end{center}

\normalsize

Properties of wave excitations in condensed matter are strongly influenced 
by spatial dispersion. In solids dispersion originates from the
discreteness of real crystals. The nonlinear dynamics of lattice models 
exhibits thus many specific phenomena known as the discreteness effects  
$[1-7]$. In the long  wave limit, when the systems are considered as 
continuous, these effects have to disappear. However some of them
leave a trace in the continuum limit, and hence can be picked as a class 
of universal dispersive effects.

Naturally the same situation is observed in macroscopic discrete
systems, e.g., arrays of Josephson junctions and nonlinear transmission 
lines, systems with nonlocal interactions, and others $[8-15]$.

The universal effects can also manifest themselves in strongly dispersive
media, in which the dispersion is not a consequence of translational symmetry 
of underlying structures. Examples of such systems are plasmas, fluids, 
optical and dissipative-dispersive systems $[16-23]$.  

One of these effects, occuring in the dynamics of nonlinear excitations of
a dispersive medium is the formation of bound states of solitons $[9,24]$.
The role of dispersion as a factor influencing the interaction between
well-separated solitons was discussed in a great number of works (see,
e.g., $[9,18,22-24]$). As a result, two mechanisms of a formation of soliton
bound states were found out. In the first case interactions of oscillating
soliton tails in dispersive media lead to the formation of bunches of
solitons, consisting of two or more well-distingushable humps [18,22]. In the
second case solitons coexisting with resonant radiation can form bound
states with purely solitonic asymptotics due to some kind of the radiation
interference effect $[25-27]$. In these theories the contribution of the
dispersion to the soliton interaction has to be considered as a weak 
perturbation. As a result the energy of the formed multisoliton structure 
differs slightly from the energy of a corresponding set of the 
non-interacting solitons. 

In the case of topological solitons this phenomenon of soliton bunching
was observed through numerical simulations beginning with the work $[8]$,
where, in fact, the weakly discrete sine-Gordon model was studied. Then
there were attempts $[28,29]$ at explaining the effect basing on the use
of the soliton perturbation theory applied to the continuous analogue of
the system. Authors of $[28]$ were the first who pointed out that any
dispersive terms should be added to the usual sine-Gordon model to obtain
the multisoliton steady solutions. Then Peyrard and Kruskal observed by means
of a numerical simulation the almost radiationless motion of the $4\pi
$-soliton in the highly discrete sine-Gordon system $[2]$. This effect 
occurs not only in the discrete model $[2,8,15]$ but also in the continuous
dispersive sine-Gordon equation, and a corresponding analytical solution
for the $4\pi$-soliton complex can be found exactly $[30,31]$. 

In general, topological multisolitons exist as discrete sets of solitonic
configurations with internal structures. They are well-studied in systems
with nonlocal interactions $[11,12,14]$. At last, bound solitons can also be
realized in systems of anharmonically interacting particles. Such models
are described by the Boussinesq-type equations with the high spatial
derivatives $[22,32,33]$, and properties of the bound solitons in these
dispersive systems were studied in detail in both cases of topological and
non-topological solitons $[22,32-37]$.

In the present paper we concentrate just on the case of strong 
interaction between topological solitons. This occurs when identical 
solitons are closely placed. As a result the repulsive potential of the 
solitons has to grow rapidly with decreasing the distance between them. 
This causes the applicability of the perturbation theory, basing on 
the "one-soliton" approximation to become invalid. Then one has to take 
into account the dispersion, as an additional source of strong 
interaction, and this can change the character of the interaction 
between solitons in general.

The aim of this paper is to show that solitons in a strongly dispersive 
medium possess an internal structure and their interaction depends on
intrinsic properties such as flexibility. Due to this dependence the 
potential energy turns out to be a non-monotonic function of the distance 
between solitons. As a result identical solitons can attract each other and
form a bound-soliton complex which can move without any radiation in
strongly dispersive media $[30,31]$. Thus we call the bound soliton
states with the zero and small distances between composite solitons  
the soliton complex and its "excited" states respectively. 

We present a number of dispersive models bearing such topological
soliton complexes. The models are described by nonlinear equations with
fourth and higher spatial or mixed derivatives. Solutions of the relevant
equations can be obtained numerically and analytically. We found exact
analytical solutions for two variants of the dispersive sine-Gordon (dSG)
and double sine-Gordon equations (dDSG). We aslo show that the complex,
consisting of three solitons, can be described explicitly in the
dispersive equations with the additional sixth derivative in the cases of
the sine-Gordon and triple sine-Gordon models.

We propose the classification of the "excited" states of soliton complexes,
constructing them explicitly in the framework of the double piecewise-linear
dispersive model. The two-soliton ansatz approximation is used to establish 
analytically the existence condition for the soliton complex in the dSG
and dDSG equations. We find numerical solutions of these equations for 
the two-soliton complex and its "excited" states, and their dependences of 
energies and velocities on the model parameters. As a result we formulate the 
concept of the soliton complex and classify it as a specific bound state of 
strongly interacting identical solitons in a dispersive medium.

\bigskip

\medskip

{\bf 2. DISPERSIVE MODELS WITH SOLITON COMPLEXES}

\bigskip

As a first example of a dispersive model we mention the discrete sine-Gordon 
system which is described by the equation $[2,4,38]$:

\begin{equation}
\label{lSGE}\frac{\partial ^2u_n}{\partial \tau ^2}+2u_n-u_{n-1}-u_{n+1}+\frac
1{d^2}\sin (u_n)=0 ,
\end{equation}
where $u_n$ is, e.g., the displacement of atom $n$ and $d$ is the discreteness
parameter. A stationary motion of a single $2\pi$-soliton is impossible in 
this dispersive system because of a strong radiation emitted by the moving 
$2\pi$-soliton $[4]$. At the same time numerical simulations $[2,8,15]$ 
showed the almost radiationless motion of the $4\pi$-soliton and other 
$2m\pi$-soliton complexes. Authors of $[2]$ tried to explain the formation 
of the soliton complex of two identical $2\pi$-solitons by exploiting the 
fact of the presence of the Peierls potential in the lattice under 
consideration. 

However in works $[30,31]$ it was found that the radiationless motion 
of the complex can be described explicitly in the framework of the dispersive 
sine-Gordon equation (1dSG) with a fourth spatial derivative:

\begin{equation}
\label{1dSGE}u_{tt}-u_{xx}-\beta u_{xxxx}+\sin (u)=0 .
\end{equation}
Eq. (2) is obtained as the long wave limit of Eq. (1) by substituting 
$xd$ for $n$, $td$ for $\tau $, and the second difference by the series:

\begin{equation}
\label{ser}u_{n-1}+u_{n+1}-2u_n\approx u_{xx}+\beta u_{xxxx}+...
\end{equation}
The relation between the dispersive factor $\beta$ and the discreteness 
parameter $d$ is $\beta \equiv 1/(12d^2)$. The exact solution of Eq. (2), 
corresponding to the soliton complex, has the following form: 

\begin{equation}
\label{4pi}u_{4\pi }=8\arctan \{\exp (\sqrt{\frac 23}\frac{x-V_0t}{\sqrt{%
1-V_0^2}})\} ,
\end{equation}

\begin{equation}
\label{V}V_0=\pm \sqrt{1-\sqrt{\frac{4\beta }3}} .
\end{equation}
The velocity of the complex is not an arbitrary constant in the solution (4), 
but it is a function of the dispersive parameter $\beta$. As a function of 
the discreteness parameter, it equals $V_0(d)=\pm \sqrt{1- (1/3d)}$ and 
differs less than five percents from numerical result of Peyrard and Kruskal, 
as it follows from a comparison of Fig. 1 (solid line) and Fig. 12 of $[2]$. 
This fact provides the starting point of our investigation because it shows 
that taking into account a high-order dispersion in the continuum model 
described by Eq. (2), leads to the same phenomenon as that in the discrete 
model of Eq. (1). 

From the other hand it is clear that the continuum model of Eq. (2) 
describes properly stationary moving nonlinear excitations in the original 
discrete model of Eq. (1) only in the limit $\beta\ll1$. In this case 
using Eq. (4) one can write the discrete $4\pi$-soliton in the first 
approximation as 

$$  
u_{n}\simeq8\arctan \{\exp [\sqrt{\frac2d}(n-V_0\tau)]\}.
$$
For small $\beta$ (large $d$) the solution is a smoothly-varying function 
of number $n$, and its effective width is proportional to $\sqrt d$. 
This provides a validity of the continuum approximation. It is remarkable
and surprising that the analytical expression $V_0(d)$ for a velocity holds 
good in the highly discrete system, i.e., far from the continuum limit.
It would be also noted that the $4\pi$-soliton appears to be localized more
strongly than the $2\pi$-kink of the usual sine-Gordon equation, since the
kink width is proportional $d$.

It is evident $[31]$ that solutions similar to Eq. (4) are available for the
second or "regularized" dispersive sine-Gordon equation (2dSG) which has the
fourth spatio-temporal mixed derivative instead the fourth spatial derivative
of Eq. (2):

\begin{equation}
\label{2dSGE}u_{tt}-u_{xx}-\beta u_{ttxx}+\sin (u)=0 .
\end{equation}
The form of the soliton complex solution of the equation is the same 
as for the dSG equation, however the velocity dependence on the parameter 
$\beta$ differs from Eq. (5).

\begin{equation}
\label{V1}V_{r}(\beta)=\pm (\sqrt{1+\frac \beta 3}-\sqrt{\frac \beta 3}) .
\end{equation}
As a function of the parameter $d$ the velocity is also shown in Fig. 1 
(dash line). 

The soliton complex with the internal structure was virtually identified by
Peyrard et al $[39]$ in simulations of the continuous modified sine-Gordon
model. In general, the taking into account the spatial dispersion or nonlocal
interactions leads to a possibility of the complex formation $[40]$. In fact,
the presence of freely moving soliton complexes and their "excited states" was
examined numerically in the continuous nonlocal sine-Gordon models in works
$[11,12,14]$. Such models are described by integro-differential equations
which can be reduced to systems of two local equations in the case of
exponentially-decaying kernel.  As an example, we point out the equation
$[11]$:

\begin{equation}
\label{nlSGE}u_{tt}+\sin (u)= {\frac \partial {\partial x}} 
{\int\limits_{-\infty }^\infty dx' G(x-x') u_{x'}(x',t)}  .
\end{equation} 
In the case of the kernel 
$G(x,\lambda)=(1/2\lambda)\exp(-\left|x\right|/\lambda)$ 
this equation is transformed into the set of equations

$$
\lambda^2 w_{xx}-w=-u_{x},\hspace{1cm}u_{tt}+\sin(u)=w_{x} ,
$$ 
which possess soliton-complex solutions.

All the above facts collected together prompt that the formation 
of the soliton complexes is a universal property of the strongly 
dispersive media. Therefore, one can try to find solutions for the soliton 
complexes in more general systems than the dispersive sine-Gordon models.

The present paper is mainly devoted to studying the soliton complex in the
dDSG equations because of two reasons. First, the usual DSG equations describe
a large variety of physical systems: ferro- and antiferromagnets,
magneto-elastic systems, superfluid $^{3}$He and others $[41-43]$. Secondly,
in the usual DSG equation $2\pi$-kinks form the $4\pi$-kink, the wobbler,
due to the action of an external field. In the dispersive DSG equations both
factors, dispersion and external field, cause the soliton coupling, and it is
interesting to investigate their mutual influence on the binding process.
At last the dDSG equations contain the sine-Gordon equations as the limit 
cases.

Thus we deal with the dispersive double sine-Gordon equation (1dDSG):

\begin{equation}
\label{dDSGE}u_{tt}-u_{xx}-\beta u_{xxxx}+\sin (u) +2h\sin (\frac u 2)=0 ,
\end{equation} 
and its regularized variant (2dDSG): 

\begin{equation}
\label{dRSGE}u_{tt}-u_{xx}-\beta u_{ttxx}+\sin (u) + 2h\sin (\frac u 2)=0 ,
\end{equation} 
where, for example, in magnetic applications 
$\phi (x,t) = {\frac12}u(x,t)$ 
denotes the azimuth angle of the magnetization vector in the easy-plane 
ferromagnet, and $h$ is a magnetic field applied along the easy plane. 
When $h=0$ the equations revert to the dispersive and regularized 
sine-Gordon equations. 

Eq. (9) can be derived from the Lagrangian:
\begin{equation}
\label{Lag}L=\int\limits_{-\infty }^\infty \frac 12\{u_t^2-%
u_x^2+\beta u_{xx}^2-2(1-\cos (u))-8h(1-\cos (\frac u2))\}dx .
\end{equation}
The Hamiltonian of the dispersive double sine-Gordon system is given by:
 
\begin{equation}
\label{Ha}H=\int\limits_{-\infty }^\infty \frac 12\{u_t^2+%
u_x^2-\beta u_{xx}^2+2(1-\cos (u))+8h(1-\cos (\frac u2))\}dx .
\end{equation}
Corresponding expressions for the regularized Eq. (10) are obtained 
by substituting $u_{tx}^2$ for $u_{xx}^2$ in Eqs. (11) and (12).

The difference between the dDSG equations results in two types of spectra 
of linear excitations. Namely, Eq. (9) and Eq. (10) have the dispersion 
relations $\omega (k)=\sqrt{1+h+ k^2-\beta k^4}$  and  
$\omega (k)=\sqrt{(1+h+ k^2)/(1+\beta k^2)}$ , respectively. For the 
first spectrum there exists formally the critical wave number 
$k_0$ at which $\omega (k_0)=0$ . 
This means that equilibrium state $u=0$ is unstable with respect to the short 
wave length perturbations. Recalling the spectrum of linear excitations of 
the discrete system one realizes the artificial origin of this instability.
The regularized equation is introduced in order to avoid this instability 
$[33,44]$ as easily seen from its spectrum. Moreover, such a spectrum 
shares the main features of Eq. (8), the sine-Gordon model with the 
nonlocal interaction.

Some important properties of the dSG and dDSG equations can be reproduced 
by the following model equation which we call as the dispersive double 
piecewise-linear equation:

\begin{equation}
\label{dDPE}u_{tt}-u_{xx}-\beta u_{xxxx}+ f(u)=0 ,
\end{equation} 
where force $f(u)$ is a periodic function of the period $4\pi$.
On the interval $[-2\pi,2\pi]$ it is given by   

\begin{equation}
\label{force}
f(u)=\left\{ \begin{array}{ll}
(1+h)(u+2\pi)& -2\pi<u<-u_0\\
(1-h)u&\left| u \right| < u_0\\
(1+ h)(u-2\pi)& u_0< u <  2\pi .
\end{array} \right.  
\end{equation}

The double quadric potential of the model is presented in Fig. 2. The point,
corresponding to the energy maximum, is chosen as $u_0=2\arccos(h)$ to model 
the behaviour of the double sine-Gordon model. As $h=0$ and $h=1$ the 
equation degenerates into the analogue of the dSG equation with the periods 
$2\pi$ and $4\pi$, respectively. We show further that this model allows to 
construct explicitly the expressions for the soliton complexes and their 
"excited" states.

Other dispersive equations combining the properties of the Klein-Gordon 
and Boussinesq-type equations are known $[34-36]$ , in which topological 
solitons are shown analytically and numerically to possess an internal 
structure, and, therefore, they are good candidates for bearing soliton 
complexes.

\bigskip 

\medskip

\begin{center}
{\bf 3. EXACT SOLUTIONS FOR SOLITON COMPLEXES AND THEIR DYNAMICAL PROPERTIES}
\end{center}

\bigskip

The exact moving solution of Eq. (4)  describes the soliton 
motion without radiation in a dispersive medium. This radiationless 
dynamics of a soliton complex in the dSG equation was discussed in detail 
$[30,31]$. In this section we present exact solutions for the 
dispersive double sine-Gordon equations and discuss the peculiarities 
of their dynamics. For the concrete definition we imply the magnetic 
applications of the equations, where solitons are domain walls and 
parameter $h$ is a magnetic field. Therefore we use further this 
terminology. 

It is known that applying a magnetic field to a ferromagnet leads to coupling 
$180^{o}$ domain walls into $360^{o}$ domain wall. In terms of the usual  
double sine-Gordon equation ($\beta=0$) this means the existence of a wobbler 
solution $[41,42]$ or the $4\pi$-kink. Such a static solution does not exist 
for Eq. (9) if $\beta \ne 0$ but it holds for Eq. (10): 

\begin{equation}
\label{twosow}u_{w}(x)=4\arctan \{\exp (q_wx-R_w)\}+%
4\arctan \{\exp (q_wx+R_w)\}
\end{equation}
where $q_w=\sqrt{(1+h)}$ and $\sinh(R_w)=1/\sqrt{h}$ .

In this work we are interested in a stationary soliton motion, i.e. consider 
the solutions of the form $u(x,t)=u(x-Vt)$. Then both the equations (9) and 
(10) are reduced to the ordinary differential equation:

\begin{equation}
\label{odnr}u_{zz}+\alpha u_{zzzz}-\sin (u)-2h\sin (\frac u 2)=0 .
\end{equation} 
Here $z=(x-Vt)/\sqrt{1-V^2}$, and parameter $\alpha$ equals to
$\alpha^{(1)}$ ($\alpha^{(2)}$) for the 1dDSG (2dDSG) equation, where

\begin{equation}
\label{alf}\alpha^{(1)}=\frac {\beta}{(1-V^2)^2},\hspace{1cm}\alpha^{(2)}%
={\frac {\beta V^2}{(1-V^2)^2}} .
\end{equation} 
When $h=0$, Eq. (16) is reduced to the dSG case which was  analysed
in $[30,31]$. Some results of the numerical integration of the reduced 
equation, as the specific limit of the nonlocal model (8), were 
discussed in $[11]$.

When $h \ne 0$ we are able to find again the exact solution of Eq. (16):   

\begin{equation}
\label{dDSGsol}u_{4\pi }=8\arctan \{\exp (\sqrt{\frac23+h}\frac{x-Vt}%
{\sqrt{1-V^2}})\} .
\end{equation}
Thus the 1dDSG and 2dDSG equations have complex solutions in identical 
form but with different expressions for the velocities   

\begin{equation} 
\label{vel1}V_{1}(\beta ,h)=\pm \sqrt{1-(2+ 3h)\sqrt{\frac \beta 3}} .
\end{equation}

\begin{equation} 
\label{vel2}V_{2}(\beta ,h)=\pm (\sqrt{1+\frac \beta 3(1+\frac 32h)^2}%
-\sqrt{\frac \beta 3}(1+\frac 32h)) .
\end{equation}
These solutions describe complexes consisting of two strongly bound 
$2\pi$-solitons. They differ from the wobbler Eq. (15) by  
the effective widths, the zero distance between solitons, and the ability to
move in the dispersive medium. 
Since now velocities are functions of parameter $h$, we can change them
from the maximum values corresponding to the dSG limit (Eq. (5) 
and Eq. (7)) to zero. In Eq. (9) it occurs at the finite critical value 
$h_{cr}=\sqrt{\frac 1{3\beta}}-\frac 23$. For example, in ferromagnets we 
can control the velocity of a motion of the domain wall complex through 
the magnetic field.  

The next peculiarity of the complex dynamics is revealed, if we evaluate 
the energies using the Hamiltonian expressions (see Eq. (12)). For the 
1dDSG complex we obtain:
 
\begin{equation}
\label{eneD1}E_{1}=32[(3\beta)^{-\frac 14}-{\frac 29}(3\beta)^\frac 14] .
\end{equation}
The energy of the soliton complex turns out to be independent of the
parameter $h$, and, therefore, of the velocity at the fixed $\beta$! 
This means that, at least, by abiabatic variation of the magnetic field, 
we can vary the form and speed of the complex, conserving its energy.
This remarkable property could be used in the energy transfer applications 
in physical systems described by the 1dDSG equation. However, the property 
is very much sensitive to spectrum characteristics of the dispersive medim. 
The regularized Eq. (12) has the velocity and field dependent energy:

\begin{equation}
\label{eneD2}E_{2}=32[(3\beta {V_2}^{2})^{-\frac 14}-{\frac 29}(3\beta %
{V_2}^{2})^{\frac 14}] ,
\end{equation}
where ${V_2}(\beta,h)$ is given by the expression (20). More details 
of comparative analysis of the energy dependences of the complexes are 
presented in the section 6.

The next natural question is the existence of an exact solution for soliton 
complexes consisting of more than two solitons. Numerical integrations of 
the discrete sine-Gordon equation Eq. (1) and the nonlocal model Eq. (8) 
reveal such solutions. We are able to find the analytical solution 
describing the three-soliton complex in the dispersive sine-Gordon 
equation with sixth spatial derivative $[45]$:
 
\begin{equation}
\label{3dSGE}u_{tt}-u_{xx}-\beta u_{xxxx}-\gamma u_{xxxxxx} + \sin (u)=0 .
\end{equation}
The equation can be derived from a discrete model if we take into account
the higher-order term in an expansion of the type of Eq. (3). For the special
choice of the parameters $\gamma ={\frac 3{20}}\beta^{\frac 32}$, Eq. (23) 
has the following exact solution:  
 
\begin{equation}
u_{6\pi }=12\arctan \{\exp (\sqrt{\frac {23}{45}}\frac{x-V_{*}t}%
{\sqrt{1-V_{*}^2}})\} ,
\end{equation}
where the velocity takes the form:

\begin{equation}
\label{Vs}V_{*}=\pm \sqrt{1- {\frac {23} {30}}{\sqrt\beta}} .
\end{equation}
The exact solution of the form of Eq. (24) exists also in the dispersive 
triple sine-Gordon equation $[45]$:

\begin{equation}
\label{3dDSGE}u_{tt}-u_{xx}-\beta u_{xxxx}-\gamma u_{xxxxxx}%
+ \sin (u)+ h_{1}\sin({\frac u3}) + h_{2}\sin (\frac 23 u)=0 ,
\end{equation} 
which is the generalization of Eq. (23), and where parameters $h_1$ and $h_2$ 
are arbitrary constants. Under the condition
$\gamma ={\frac 3{20}}\beta^{\frac 32}(1+\frac 12 h_2)$ the three-soliton 
complex is described by the expression:

\begin{equation}
\label{resh}u_{6\pi }=12\arctan\{\exp(q_{3}(x-V_{3}t)\} .
\end{equation}  
The relations between solution parameters are the following:

\begin{equation}
\label{q_resh}q_3=(\frac {4+2h_2}{9\beta})^{1/4} ,
\end{equation}

\begin{equation}
\label{V_resh}V_{3}= \pm \sqrt{1-({\frac {23}{15}}+h_1-%
{\frac 83h_2})\sqrt{\frac {\beta}{4+2h_2}}} .
\end{equation}
In previous formulas for the presentation of exact solutions we have used the
Lorentz-invariant-like expressions to keep the analogy with those of the
Lorentz-invariant equations with $\beta=0$. However the solutions can be 
written also in the form like Eq. (27) which is evidently simpler. 
In any case at a first glance the solutions for the soliton complexes and 
conditions of their existence look like exotic.   
This raises, at least, three questions. The first of these: how are the 
solutions sensitive to the variation of the equation parameters? The second, 
what kind of other solutions exists in the strongly dispersive equations? 
And the third, what is the mechanism underlying the creation of these 
bound states? Pondering these questions provides the substance of the 
next sections.

\bigskip

\medskip
\begin{center}
{\bf 4. PHENOMENOLOGY OF SOLITON COMPLEX}
\end{center}

\begin{center}
{\bf A. Collective  coordinate approach to soliton-complex formation}
\end{center}

\medskip

An analytical approach to the description of the soliton-complex formation 
in the dSG equation was proposed in $[30,31]$. It is based on the use of 
the collective variable ansatz which is constructed by taking into account 
the translational and internal degrees of freedom of a soliton as well 
as interactions between solitons and solitons with radiation:

\begin{equation}
\label{Anz}u(x,t)=u^{(s)}(x,t;l,X,R)+u^{(r)}(x,t) .
\end{equation}
Here $u^{(s)}$ is a solitonic part and $u^{(r)}(x,t)$ is a part of the
solution corresponding to radiation. It turns out that the condition 
of the complex formation of the closely sited solitons can be found from 
the energy expression of the pair of strongly interacting solitons 
without taking into account the radiation $[31]$. Now we use this 
approximation for the description of the dispersive double sine-Gordon 
system. So we suppose that the complex dynamics can be considered in 
the framework of the soliton ansatz: 

\begin{equation}
\label{twoso}u^{(s)}=4\arctan \{\exp ({\frac {x-X}{l}}-R)\}+4\arctan \{\exp
({\frac {x-X}{l}}+R)\} ,
\end{equation}
which is prompted by the forms of the wobbler Eq. (15) and the exact solution 
in Eq. (18). Here $l=l(t)$, $X=X(t)$ and $R=R(t)$ are functions of time. 
Functions $l(t)$ and $X(t)$ describe the changing of the effective width
of solitons and their translational motion, respectively. The function 
$R=R(t)$ corresponds to the changing separation between solitons, which is 
defined obviously as $L=2lR$.

Inserting the ansatz into Eq. (11) and Eq. (12) we find the effective 
Lagrangian and Hamiltonian of two interacting solitons in strongly 
dispersive media.

\begin{equation} 
\label{lagham}L_{eff}=T-U,\hspace{1cm}H_{eff}=T+U ,
\end{equation}
where $T$ and $U$, the kinetic and potential energies, are given as

\begin{equation}
\label{kin0}
\begin{array}{c}
T={\frac {\D 8}{\D l}}\{{\frac {\D 1}{\D 3}}{l_t}^2[2{R^2}+(R^2+%
{\frac{\D\pi^2}{\D 4}})I_{1}(R)]+%
(l{R_t}\tanh(R))^2I_{2}(R)+2R{R_t}ll_t\\%
+{X_t}^{2}I_{1}(R)\} ,
\end{array}
\end{equation}
and

\begin{equation}
\label{pot}
\begin{array}{c}
U={\frac {\D 8}{\D l}}\{I_{1}(R)+l^2I_{2}(R)+ 2hl^{2}I_{0}(R)\\-%
{\frac{\D \beta}{\D l^2}}[{\frac {\D 1}{\D 3}}-{\tanh^2(R)}(1-%
I_{2}(R)(1+{\frac{\D 2}{\D \sinh^2(2R)}}))]\} ,
\end{array}
\end{equation}
where the following notations have been introduced (see also Appendix A):
$$
\begin{array}{c}
I_{0}(R)=2R\coth(R) ,\\
I_{1}(R)=1+\frac { \D 2R}{\D \sinh(2R)} ,\\
I_{2}(R)={\coth^2(R)}(1-\frac {\D 2R}{\D \sinh(2R)}).
\end{array}
$$
Details of the calculation of explicit expressions for the kinetic and
potential energies are presented in Appendix A. The expressions look 
rather complicated, but they contain a rich information about two-soliton 
dynamics and interactions. In particular, in the limit cases they 
describe the exact two-soliton solution of the integrable sine-Gordon 
equation and the small wobbler oscillations in the usual double 
sine-Gordon equation (see Appendix B and C).

Since we are interested in the stationary soliton complexes let us
investigate at first the potential energy of two interacting solitons. 
The energy is drawn in Figs. 3(a)-3(d) as a function of parameters $l$ 
and $R$ for two different values of the dispersive parameter $\beta$.
For the sake of symmetry we show the energy dependence on both 
positive and negative values of $R$, because the ansatz (31) and the 
energy are even functions of $R$.

Let us examine the character of the dependence of the 
potential energy on the separation between solitons at the various $l$. 
In Figs. 3(a) and 3(b) we take $\beta=\beta_1={\frac 15}$. One can see 
that in the whole area shown the closely placed solitons repulse each 
other and the energy has a maximum at  $R=0$. There exists the critical 
effective length $l_{cr}$ at which two local minima appear for the first 
time at  $R=\infty$, and then, when $l$ increases, they quickly diminish 
their separation. The minima correspond to the equilibrium positions of 
two solitons in which their mutual repulsion is balanced by the opposite 
action of the dispersive part of the interaction and the magnetic field. 
We show two cases, when $h=0$ and $h=0.1$ with a view to compare results 
for the sine-Gordon and double sine-Gordon systems. Fig. 3(b) demonstrates 
the strong attractive contribution of the magnetic field to the soliton 
interaction. 

In Figs. 3(c) and 3(d) the parameter $\beta$ is chosen as $\beta_0=\frac 34$.
One can find that in this case the critical value $l_{cr}$ corresponds the
point where the local maximum in the $R$-dependence is changed by a local 
minimum. This obviously means that in this point the repulsion between 
solitons is changed into the attraction. 
This fact can be proved strictly and the critical values can determined
exactly because at small $R$ the consideration can be performed 
analytically (see Appendix A for details).
Indeed, from the Largangian expression Eqs. (87-89) valid for small $R$ one 
derives the following Langrange equations:

\begin{equation}
\label{Leq2}{\frac{\pi ^2}6}ll_{tt}-{\frac{\pi ^2}{12}}l_{t}^2%
-1+X_{t}^2+({\frac 1 3}+h){l^2}+{\frac{\beta}{l^2}}=0 ,
\end{equation}

\begin{equation}
\label{Leq1}{\frac {X_{t}} l}=p=const ,
\end{equation}

\begin{equation}
\label{eq3}X_{t}^2 -1+({\frac 1 5}+h)l^2+{\frac75}{\frac{\beta}{l^3}}=0 .
\end{equation}
The stationary values of  $l_0$ and $X_{t}=V_0$ are easily found. They    
coincide with parameters of the exact solution for the soliton complex, 
Eqs. (18) and (19), and give the critical  values for the parameters at 
which the soliton attraction arises. It is easy to see, when Eq. (37) is 
satisfied, that the contribution of terms of $O(R^2)$ to the "stationary" 
Lagrangian part Eq. (89) equals zero, i.e. at this point it changes sign.   
That is, for example, for $V_0=0$ and $h=0$, one finds from Eqs. (19) 
the following critical parameter values: $\beta_0=\frac 34$ and 
$l_{0}(\beta_0)=l_{cr}=\sqrt{3/2}\approx1.225$. 
In fact, at this value $l_{cr}$ the potential energy exhibits the flat 
plot at small $R$ as it is seen in Fig. 3(c).  
Hence the possibility of solitons to change their effective length in the
dispersive medium leads to changing the very character of interaction between
them. Thus the exact complex solutions correspond to the bound states of
solitons coupled in reason of their own attraction.

Stability of the complex with respect to changing its parameters, velocity and
effective width, as well as a possibility of its decay can be tested by the
numerical simulation of the equations. This work is in progress. Results of
$[2,11,15]$ give information about the stability of the complexes in the 
discrete and nonlocal sine-Gordon model. The complexes manifest themselves as
attractors in the soliton dynamics of these dispersive media. Two solitons,
moving with velocities larger than the critical value radiate energy until the
velocity reaches its stationary value. Then the formed complex moves
radiationlessly.  At a velocity smaller than the critical value solitons
repulse each other, and the complex decays. As a result the composite solitons
travel from the center to their new equilibrium positions. 

This picture is consistent with conclusions following from our direct energy 
analysis. Further consideration shows that, for the first time,
equilibrium local minima at $R=\infty$ appear for the following values of
the parameter $l$: 

\begin{equation}
\label{lbeta}l(\beta)= \sqrt{\{[{\frac14}+\beta(1+h)]^{1/2}+%
{\frac12}\}/(1+h)} .
\end{equation}   
It is easy to see that the parameter $l_{0}$ of the exact complex solution
also belongs to this dependence. Therefore at the same moment , when the
attraction between solitons arises, the equilibrium local minima disappear 
and vice versa. 
The presence of the local minima points out on a possibility of the existence
of another attractor of the wobbler-like type in soliton dynamics. 
To study the question about the existence of other stationary moving
bound states of identical solitons, besides the exact soliton complex, 
we recur to the ordinary differential equation (16). 
 
\begin{center}
{\bf B. Excited states of the soliton complex}
\end{center}

The equation (16) for stationary states has the asymptotics $u(z)=A\exp(qz)$
where $q$ obeys the equation 

\begin{equation}
\label{disp1}\alpha q^4 + q^2 -1-h=0 .
\end{equation} 
Soliton solutions are characterized by the vanishing asymptotics, hence the 
corresponding $q=\kappa$ is supposed to be a real parameter:

\begin{equation}
\label{kap}\kappa=\sqrt{\{[{\frac14}+\alpha (1+h)]^{1/2}-%
{\frac12}\}/ \alpha}. 
\end{equation}   
The exact complex solution is realized when $\alpha=3/4$.

However Eq. (16) is of the fourth order and has another asymptotics which is
oscillating one. In this case $q=ik$, where

\begin{equation}
\label{k}k= \sqrt{\{[{\frac14}+\alpha(1+h)]^{1/2}+{\frac12}\}/ \alpha} .
\end{equation} 
In general, a solution of Eq. (16) includes the solitonic and
oscillating parts. This is simply interpreted, because a single moving 
$2\pi$-soliton in the dispersive system usually radiates energy and 
generates the continuous waves. Thus in the case of the moving soliton we 
deal with the self-modulated medium. It is interesting to note that it 
occurs independently of the stability property of the equilibrium state. 
In fact, both equations (9) and (10) are reduced to Eq. (16), 
in spite of different dispersion relations.   

When two solitons are present in the system, they create the radiation
background, and one soliton moves upon the undulations generated by the other
soliton. The situation turns out to be similar to that in the original
discrete system $[2]$, where solitons travel upon the Peierls potential. The
existence of the purely solitonic stationary excitations under such conditions
implies that some interference effect takes place, which leads to cancelling
the radiation far from soliton complex. Available theories [25-27] of a
soliton binding by radiation field suggest a large separation between
composite solitons and a small influence of the dispersion on soliton
interactions. It is not correct for closely-sited solitons forming the soliton
complex. As a result an alternative approach to the description of these bound
states is required. 

To find all possible forms of the soliton-complex solutions one has to 
solve the nonlinear eigenvalue problem, such as Eq. (16), which is hardly 
feasible by analytical tools. There are known some attempts to solve 
similar problems by variational methods $[26,46]$. We propose another 
approach to a searching for soliton-complex solutions and demonstrate it 
by applying to Eq. (16). 

So we seek solutions of Eq. (16), imposing the vanishing boundary 
conditions. Let us reformulate this problem as a self-consistent eigenvalue 
problem for the linear equation:

\begin{equation}
\label{shr1}[-\alpha {\frac {\D d^2}{\D dz^2}} +U(z)]u_{zz}=0 ,
\end{equation} 
where the potential well is 

\begin{equation}
\label{op2}U(z)=\frac{\sin (u)+2h\sin (u/2)}{u_{zz}}-1 .
\end{equation}
We know that Eqs. (42) and (43) have, at least, the solution
corresponding to the exact soliton complex for $\alpha=3/4$. To
investigate the possible existence of other forms of soliton 
complexes, let us insert the soliton anzats (31) with $R=0$: 

\begin{equation}
\label{anz2}u_0(z)=8\arctan (\exp (z/l)) ,
\end{equation}
into the potential expression Eq. (43). Then the linear equation (42) 
takes the form

\begin{equation}
\label{shro1}[-\alpha \frac{d^2}{dz^2}-1+l^2(1+h-\frac 2{\cosh
{}^2(z/l)})]\psi (z)=0 .
\end{equation}
After introducing the new coordinate $y=z/l$ Eq. (45) can be 
rewritten as 

\begin{equation}
\label{shro2}[-\alpha l^{-4}\frac{d^2}{dy^2}+1+h-l^{-2}-\frac 2{\cosh
{}^2(y)}]\psi (y)=0 .
\end{equation}
Solutions of Eq. (46) give us the next approximation for $u_{zz}$.  

Recalling asymptotics of the eigenvalue problem and Eq. (40), we 
have to impose the following condition on parameters $\alpha$ and $l$:

\begin{equation}
\label{lev1}\alpha l^{-4}=1+h-l^{-2}\equiv \epsilon  ,
\end{equation}
which must hold for any soliton complex.
We have introduced the parameter $\epsilon$ which takes, evidently, a 
discrete set of values $ \epsilon _n$ . Thus finally we find the equation:

\begin{equation}
\label{shro3}[-\frac{d^2}{dy^2}+1-\frac{2\epsilon ^{-1} _n}{\cosh {}^2(y)}%
]\psi _{n} (y)=0 .
\end{equation}
Eigenvalues and eigenfunctions of this type of the equation are 
well-known $[47]$. In particular, the equation for parameter 
$ \epsilon _n$ reads:

\begin{equation}
\label{lev0}1=\frac 14[(1+8\epsilon {}^{-1}_n)^{1/2}-2n-1]^2 .
\end{equation}
As a matter of fact, the problem is now reduced to the determination of 
all values of the potential depth in Eq. (48) for which the discrete 
level equals unity. From Eq. (49) it follows

\begin{equation}
\label{lev2}\epsilon _n=2/[(n+1)(n+2)] ,
\end{equation}
where $n=1,2,3...$. Then one obtains for $l_n$ and $\alpha _n$ 

\begin{equation}
\label{ln1}l_n^{-2}=h+\sigma _n ,
\end{equation}
where $\sigma_n$ equals

\begin{equation}
\label{An}\sigma _n=\frac{n(n+3)}{(n+1)(n+2)} ,
\end{equation}
and

\begin{equation}
\label{alf01}\alpha _n=\alpha _n^0(1+\frac h{\sigma _n})^{-2} .
\end{equation}
Here $\alpha _n^0$ are eigenvalues of the problem for the case  
$h=0$, i.e. for the dispersive sine-Gordon problem:

\begin{equation}
\label{alf0}\alpha _n^0=\frac{2(n+1)(n+2)}{[n(n+3)]^2} .
\end{equation}
Ten first eigenvalues $\alpha _n^0$ are shown as solid circles in Fig. 4
(to show more vividly the dependence on $n$ we connect the points by 
the solid line). The infinite series of discrete values of $\alpha _n^0$ 
rapidly diminishes with increasing $n$. Only odd values of $n$ seem 
to be valid for the problem under the consideration. However, as  
shown in Fig. 4, the even values of $n$ also reproduce well enough the 
exact eigenvalues (solid squares) which we find by numerical 
integration of Eq. (16) (see section 6). This is explained by the fact 
that the next iteration step leads to splitting of levels with 
$n > 1$ into two close levels with even and odd eigenfunctions.  
The level splitting is smaller for higher $n$, and the analytical 
dependence $\alpha _n^0$ serves as a good approximation of the eigenvalues 
in the dispersive sine-Gordon case. 

Eigenfunctions of Eq. (48) also describe well the changing of the form of 
$4\pi$-soliton when it appears in the excited state. Functions $\psi _n$, 
corresponding to $u_{zz}$, vanish exponentially at $z=\pm\infty$ and 
exhibit an oscillating behavior at the coordinate origin.  The oscillation 
domain increases for higher $n$-values. This oscillation can be interpreted 
as the radiation locked between two $2\pi$-solitons. Thus the soliton 
complex can be realized by the infinite series of configurations to be 
referred to, naturally, as "excited" states.  
 
Figure 5 presents $\alpha _n$ of five levels as functions of the parameter 
$h$. We see that values of $\alpha _n$ tend quickly to zero at high fields, 
reflecting correctly the qualitative tendency in the $h$-dependent behavior 
of exact eigenvalues. Quantitative comparison will be done in section 6. 

\bigskip

\begin{center}
{\bf 5. EXACT SOLITON COMPLEX SOLUTIONS IN DOUBLE QUADRIC MODEL}
\end{center}

Explicit expressions for $4\pi$-soliton complex and its "excited" states 
can be found in the framework of the double piecewise-linear model Eqs. (13) 
and (14) $[48]$. In this case stationary moving complexes can be 
constructed in analytical form , and hence, the corresponding 
eigenvalue problem for the parameter $\alpha_n$ can be exactly solved. 

In this section we exhibit principal results of the considerations. 
For the sake of simplicity we present main formulas in the limit case 
$h=0$ in reason of their clarity. 

As in the dDSG case, after introducing the coordinate $z$ in the moving  
reference frame, we derive the equation:

\begin{equation}
\label{DD1}u_{zz}+\alpha u_{zzzz}-f(u)=0 ,
\end{equation}
where $f(u)$ is given by the expression Eq. (14). We are interested in 
odd solutions of Eq. (55), $u(-z)=-u(z)$, with limit conditions 
$u(\pm\infty)=\pm 2\pi$.
To construct the $4\pi$-soliton it is sufficient to find the solution in
the external region ($z>z_0$) and the general odd solution in the internal 
region ($\left| z \right|<u_0$). They look like the followings: 

\begin{equation}
\label{DD3}u_e(z)=2\pi -A\exp (-\kappa _2z) ,
\end{equation}
and

\begin{equation}
\label{DD2}u_i(z)=B\sin (k_1z)+C\sinh (\kappa _1z) ,
\end{equation}
In the limit case $h=0$ the exponent $\kappa _2$ coincides with the 
parameter $\kappa _1$ of the solution $(57)$: 

\begin{equation}
\label{D15}\kappa _1=\kappa _2=\kappa =\pm \sqrt{[(\frac 14+\alpha%
)^{1/2}-\frac 12]/\alpha } .
\end{equation}
The parameter $k_1$ equals to: 

\begin{equation}
\label{D16}k_1=k=\sqrt{[(\frac 14+\alpha )^{1/2}+\frac 12]/\alpha } .
\end{equation}
Using the conditions of continuity of the function $u$ and its 
first, second and third derivatives in the point $z_0$, where $u=u_0$,
we arrive at the following set of equations:

\begin{equation}
\label{D04}B\sin (kz_0)+C\sinh (\kappa z_0)=2\pi -A\exp (-\kappa z_0)=u_0 ,
\end{equation}

\begin{equation}
\label{D05}kB\cos (kz_0)+\kappa C\cosh (\kappa z_0)=A\kappa \exp
(-\kappa z_0) ,
\end{equation}

\begin{equation}
\label{D06}-k^2B\sin (kz_0)+\kappa ^2C\sinh (\kappa z_0)=-A\kappa ^2%
\exp (-\kappa z_0) ,
\end{equation}

\begin{equation}
\label{D07}-k^3B\cos (kz_0)+\kappa ^3C\cosh (\kappa z_0)=A\kappa ^3%
\exp (-\kappa z_0) .
\end{equation}
From the system Eqs. (60-63) one can find the eigenvalues of the 
parameter $\alpha$ and the coefficients $A,B$ and $C$. 

It is easy to see from the Eqs. (61) and (63) that $\cos(kz_0)=0$, and 
hence $z_0$ is given by 

\begin{equation}
\label{D17}z_0=\frac \pi k(n-\frac 12)\equiv \mu _n/k .
\end{equation}
Introducing the parameter $\lambda (\alpha)=\kappa (\alpha) /k(\alpha)$ 
we obtain the following equation: 

\begin{equation}
\label{D118}\lambda = \exp (-\lambda \mu_n) ,
\end{equation}
which determines the eigenvalues $\lambda _n$ and $\alpha _n$.
It is clear that this equation has a solution for every $n$. Moreover
it is reduced by the substitution $\Lambda _n=\lambda \mu _n$ to the 
Lambert's equation:

\begin{equation}
\label{D18}\Lambda _n \exp (\Lambda _n) =\mu_n .
\end{equation}
Its solution is the Lambert's function $\Lambda _n(\alpha) = W(\mu _n)$.  
By solving the last equation with respect to $\alpha$ we find that
there is a infinite series of the $\alpha _n$-values for which we can 
construct the $4\pi$-soliton. Coefficients of the solution are expressed 
through $k(\alpha _n)$ and $\kappa(\alpha _n)$ as follows: 

\begin{equation}
\label{D19}A=\pi \frac k\kappa ,
\end{equation}

\begin{equation}
\label{D20}B=(-1)^{n-1}\frac{2\pi \kappa ^2}{k^2+\kappa ^2} ,
\end{equation}
and

\begin{equation}
\label{D21}C=\frac{2\pi k\kappa }{k^2+\kappa ^2} .
\end{equation}

Ten eigenvalues $\alpha _n$ are shown as the open circles in Fig. 4.
Note that parameters $\alpha _n$ diminish with increasing $n$ in the like 
manner as in the analytical results Eq. (54). 

In the general case $h \ne 0$ the equation
for $\alpha _n(h)$ is much more complicated $[48]$ than Eq. (65). The 
results of its solving are presented in Fig. 6. One can see that 
$\alpha _n(h)$ are quickly decaying functions of $h$. 

Now we discuss properties of the eigenfunctions. Evidently, the soliton 
complex states can be classified by the integer values of the parameter $n$.
In the "excited" complexes the value of $n$ shows the number of nodes of the
oscillating part of the solution. The latter corresponds to the 
radiation locked between two composite solitons.

The first five eigenfunctions are shown in Fig. 7, where they are numbered
from the left to the right by $n=1,2,3,4,5$, respectively. 
For convenience of observation we have shifted the centers of complexes 
in space. With increasing $n$ the separation $L$ between the $2\pi$-solitons 
slowly grows. For large $n$, $L\sim 2z_0\simeq2\ln(\pi n/\ln(n))$. 
At the same time the amplitude of the oscillations decreases, and in 
the limit $n\rightarrow\infty$ the soliton complex is approximated by 
two well-separated solitons and a linear standing wave between them. 

It should also be noted that $u_z\geq0$ for all functions. 
More precisely, at the soliton center $z_c$ the first derivative reaches 
its local maximum value for an odd $n$ and $u_z=0$ for the even $n$. 
The latter is possible if the expansion of $u(z)$ at the point $z_c$ 
begins with a term of the order $(z-z_c)^3$. 
Indeed, it is easy to be convinced that the condition $kB+\kappa C=0$ is 
fulfilled for the solution (56), (57) with the even $n$. The same 
behavior of the eigenfunctions was found while integrating the dispersive 
sine-Gordon equation $[11]$. For the general case $h\ne0$ the condition 
$u_z(z_c)=0$ does not hold in both the piecewise-linear and dDSG models, and 
the first derivative becomes positive at the soliton center for all $n$.

Thus the simple piecewise-linear model exhibits main peculiarities of soliton
complex structure and its stationary dynamics, which turn out to be
universal for the dispersive nonlinear models.

\bigskip

\begin{center}
{\bf 6. SOLITON COMPLEXES IN THE DISPERSIVE DOUBLE SINE-GORDON EQUATION}
\end{center}

\bigskip

In this section we present results of the numerical intergation of the 
dDSG equation in the case of the stationary complex motion. We start 
with Eq. (16) and seek its $4\pi-$soliton solution, i.e. impose the 
limit conditions:

\begin{equation}
\label{D61}u(-\infty)=0,  u(\infty)=4\pi, 
u_z(\pm\infty)=u_{zz}(\pm\infty)=u_{zzz}(\pm\infty)=0 .
\end{equation} 

However, at first, we propose one more effective approach to solving 
the equation. It allows us to simplify the integration procedure and  
formulate some strict assertions about a possibility of different 
forms of the soliton complexes.

Eq. (16) describes the effective particle dynamics in the 
four-dimensional space $\{u, u_z, u_{zz}, u_{zzz} \}$. 
It has the first integral:

\begin{equation}
\label{d62}I=\alpha [(u_z^2)_{zz}-3u_{zz}^2]+u_z^2-%
2(1-\cos (u))-8h(1-\cos(\frac u2)) .
\end{equation}
By virtue of the limit conditions, Eq.(70), $I=0$ in the case of the 
soliton complex solution. In general, the phase space is determined by 
four variables. However if we found the solution $u(z)$, we could obtain 
the trajectory in the two-dimensional phase space, $\{u,u_z\}$ by 
expressing $u_z$ as a function of $u$.

Let us introduce the function $u_z^2=F(u)$ or $u_z=\sqrt{F}$. 
By the differentiation of the definition we get that $u_{zz}=\frac 12F_u$.
After substitution of these expressions to Eq. (71) we find that the 
function $F$ obeys the following second-order differential equation:

\begin{equation}
\label{d64}F+\alpha (FF_{uu}-\frac 14F_u^2)=2(1-\cos (u))+%
8h(1-\cos(\frac u2)) .
\end{equation}
From the conditions of Eq. (70) it follows that $F(u)$ must be periodic 
with the period $4\pi$ and $F_u=0$ for $u=0, 4\pi,...$. It is easy to 
find the function expansion at $u=0$:

\begin{equation}
\label{d63}F=\kappa ^2u^2+O(u^4)+... ,
\end{equation}
where $\kappa$ is given by Eq. (40).

To this end we are able to perform the numerical integration of the
equation (72). This is achieved by use of a fourth-order Runge-Kutta
method. To study the dependence of the soliton complex form on the 
parameter $h$ we begin the integration, at first, for $h=0$ (the dSG 
case). The corresponding results are presented in Figs. 4, 9, 10(a), 10(b). 
In this case some results can be also verified and compared with 
those found in the paper $[11]$. 

First of all, it is easy to be convinced that there is no $2\pi$-kink 
solution in Eqs. (72) and (16). 
Next, there are only discrete set of $\alpha _n$ 
for which the $4\pi$-soliton complexes exist. We show the first ten 
$\alpha _n$ as the solid squares in Fig. 4 and see that they are in a 
good quantitative agreement with the analytical results (54) (Fig. 4, 
solid circles). In Figs. 10(a) and 10(b) the phase portraits are drawn 
in the plane $(u, u_z)$. 
The main soliton complex trajectory is realized for 
$\alpha _1=\frac 34$ and has only one maximum. Other phase portraits 
differ by the number of maxima. Call this $n$. So it is convenient to 
classify these states by the integer $n$ using the direct analogy with
the results of phenomenological approach and the piecewise-linear model.  
 
In particular, as in the case of the piecewise-linear model all the odd
states (Fig. 10(a)) have the nonzero first derivative at the soliton center 
where $u=2\pi$. The functions corresponding to even $n$ (Fig. 10(b)) 
demonstrate another behavior. Their first derivatives at the point 
$u=2\pi$ behave as $u_z\sim(u-2\pi )^{2/3}$. The reason of the specific 
dependence is the same as that in the piecewise-linear model, namely, it 
turns out that $u(z)-2\pi\sim(z-z_c)^3$ and hence $u_z\sim (z-z_c)^2$. 
As a result $u_z=0$ in the point $u=2\pi $ for states with the even $n$. 
However as we show below, the inclusion of the magnetic field, i.e. 
taking into account the term with nonzero $h$, removes this degeneracy.

Now the functions $u_n(z)$ for the complex and its "excited" states can 
be found by numerical integration of the equation 
$u_z=\sqrt{F(u)}$. First five soliton complex states are presented 
in Fig. 9. They virtually have the same shapes as solutions of the 
piecewise-linear model (c.f. Fig. 7).

When $h\ne0$ the new mechanism of the soliton attraction begins to work. 
The magnetic field draws together the composite solitons. Results of the 
numerical solution of the Eq. (72) for this case are shown in Figs. 11, 
12(a) and 12(b). One can see that the phase portraits of all the "excited" 
states become similar with increasing $n$ even at small $h$. This means 
that in this case the form of the "excited" complex approaches the wobbler 
solution. It is clearly seen from Figs. 12(a) and 12(b), where solutions 
for $h=0.1$ are presented. It also confirmed by the analysis, performed 
below, of the dynamical characteristics of the complex and its "excited" 
states. 

Now, we discuss the dependence of $\alpha _n$ on the parameter $h$. 
For the $n=1$ there are exact soliton-complex solution of the dDSG equation
and the analytical expression for $\alpha_1(h)$ is given by: 

\begin{equation}
\label{alf001}\alpha _1(h)=\frac34 (1+\frac 32 h)^{-2} .
\end{equation}
For the "excited" complex states the dependences $\alpha_n(h)$ are found 
numerically, and the first five of them are presented in Fig. 8. They 
are qualitatively similar to those of the piecewise-linear model 
(Fig. 6) and of analytical results (Eqs. (52)-(54) and Fig. 5). Quantitative 
comparison reveals that with increasing $h$ numerical eigenvalues 
$\alpha_n(h)$ vanish more rapidly than the analytical dependences. 
However we believe that the decaying of the functions at high $h$ 
is proportional to $O(1/h^2)$. We have fitted the data for 
$\alpha_n(h)$ by the expressions of the form Eq. (53)

\begin{equation}
\label{alf02}\alpha _n=\alpha _n^0(1+\frac h{D_n})^{-2} ,
\end{equation}
where $D_n$ is the only fitting parameter. As a result we have found 
a good coincidence between the data and the analytical approximation.
For example, in the case $n=2$, the value $D_2=2/5$ provides a 
deviation less than some percents on the interval $0\le h\le 2$.
In general, $D_n$ decreases quickly with the growth of the number $n$.

At high $h$ the soliton complex and wobbler shapes have to 
degenerate to the kink of the sine-Gordon equation for the variable
$\phi=u/2$. Indeed, introducing this new variable one can exactly 
rewrite Eq. (16) as the following form of the stationary dispersive 
double sine-Gordon equation:

\begin{equation}
\label{nodnr}\phi _{\xi\xi}+\tilde{\alpha}\phi _{\xi\xi\xi\xi}-%
\sin (\phi)-\tilde{h}\sin(\phi) \cos(\phi)=0 ,
\end{equation} 
where the following notations are introduced: $\xi=\sqrt{h}z$, 
$\tilde{h}=1/h$, and $\tilde{\alpha}=\alpha h$. 
When $\tilde{\alpha}\equiv0$, Eq. (76) has the wobbler solution which 
is reduced to $2\pi$-kink in the limit $\tilde{h}\rightarrow 0$.
The soliton complex and its "excited" states exist at $\tilde{h}\ne0$ 
and $\tilde{\alpha}\ne0$. As it follows from Eq. (75), at high $h$ 
the linear dependence between parameters $\tilde{\alpha}$ and 
$\tilde{h}$ is suggested, i.e. 

\begin{equation}
\label{alf03}\tilde{\alpha}_n \approx \alpha{}^0_n D{}^2_n \tilde{h} ,
\end{equation}
In particular, in the case of the exact solution we find from 
Eq. (74) that 

\begin{equation}
\label{78}\tilde{\alpha}_1(\tilde{h})=\frac{\tilde{h}}3 (1+%
\frac23 \tilde{h})^{-2} , 
\end{equation}
and, at small $\tilde{h}$, in fact, $\tilde{\alpha}_1\approx\tilde{h}/3$. 
The eigenfunction $\phi_1(\xi)$ would be considered as the direct 
continuation of $2\pi$-kink solution of the SG equation 
for the case $\tilde{h}\ne0$. 

The eigenfunctions $\phi_n(\xi)$ for $n\ge2$ behave like the wobbler 
solution. To understand the proximity of the "excited" states and the 
wobbler we apply the iteration procedure (see section 4.B) to Eq.(76), 
starting with the wobbler-like ansatz (31):

\begin{equation}
\label{twosov}\phi(\xi)=2\arctan \{\exp (q\xi-R)\}+%
2\arctan \{\exp (q\xi+R)\} ,
\end{equation}
where now $q$ and $R$ are the constant parameters of the solution.
The ansatz seems to be advantageous over the previous one, Eq. (44),
because it includes two parameters. However, one must keep in mind that 
already after a first iteration step all parameters get definite 
values, and the eigenfunction form will be corrected after every step. 
Here we use the appropriate choice of the ansatz to obtain the 
analytical estimation for the soliton separation $R$ in the complex 
at high field $h$.  

Omitting details of the calculation we derive finally the following
equation for the determination of $\tilde{\alpha}_n(\tilde{h})$:

\begin{equation}
\label{shro4}[-\tilde{\alpha}q^4\frac{d^2}{d\xi^2}+1+\tilde{h}-%
q^2-\frac{2(\tilde{h}-\sinh^2(R))}{\cosh^2(\xi)-\sinh^2(R)}%
]\phi _{\xi\xi}=0 .
\end{equation}
From this equation, in the case of small $\tilde{h}$, we find that  
the quantity $U_0=2(\tilde{h}-\sinh^2(R))/\tilde{\alpha}q^4$ 
has to be equal to $(n+1)(n+2)$. It is easy to see that this relation 
is in a qualitative agreement with the assumption about the behavior 
of $\tilde{\alpha}_n$ (see Eq. (77)). Then we can use Eq. (77) and 
put $q\approx1$ to obtain the following relation for the determination 
of the parameter $R$:
 
\begin{equation}
\label{R}\sinh^2(R)\approx \tilde{h}(1-D{}^2_n) .
\end{equation}
As it follows from Eq. (81), when $n$ increases, the fast decay of 
values $D_n$ causes the parameter $R$ rapidly approaches the wobbler 
value $R_w$ (see Eq.(15)), even at small $\tilde{h}$.

Thus, the smallness of $\alpha_n(h)$ and its rapidly-vanishing 
dependence on $h$ and $n$, lead to the degeneracy of the complex 
"excited" states to the wobbler-like solution.

This fact can be demonstrated perfectly, if one calculates the energies of 
the soliton complex and its "excited" states as functions of the parameter 
$h$ (see Fig. 13). We have normalized these quantities by the energy of 
the wobbler. As clearly seen from Fig. 13, energies of the "excited" 
states approach the wobbler energy very quickly with increasing $h$, 
while the exact solution remains well-separated from the wobbler up 
to high $h$. 

Another form of the dDSG equation (76) can be useful to analyze 
the hierarchies of the bifurcation values of parameters $\alpha_m(h)$ 
for other multisoliton complex solutions. One can note that there is
an infinite series of solutions of Eq. (76) for $\tilde{h}=0$ and 
$\tilde{\alpha}=\tilde{\alpha}{}^0_m$, corresponding to the 
$4\pi$-complex for the variable $\phi$. When $\tilde{h}\ne 0$ these 
solutions are modified, giving new branches  
$\tilde{\alpha}_m(\tilde{h})$. It is evident that the eigenvalues
correspond to the $8\pi$-soliton complex and its "excited" states for the
variable $u(z)$ obeying Eq. (16). It is clear that at high $h$ the 
behavior of eigenvalues $\alpha_m(h)$ is approximated by 
$\tilde{\alpha}{}^0_m /h$. This is a qualitatively different behavior 
than that in the case of the $4\pi$-complex. At small $h$ the parameters 
$\alpha_m(h)$ reach their constant bifurcation values $\alpha_m(0)$, 
which can be found from the numerical solution of Eqs. (16) and (71).

It is important to note that a concrete physical system is characterized 
by the definite value of the dispersive parameter $\beta$. Then for the 
given $\beta$ and $h$ several soliton complex states can exist 
simultaneously. 

In the case of the 1dDSG equation at the fixed $\beta$ there exists a 
discrete set of velocity values $V_n(h)$ corresponding to the 
stationary radiationless motion of the complex:

\begin{equation} 
\label{82}V_n(h)=\sqrt{1-(\beta /\alpha _n(h))^{1/2}} ,
\end{equation}
where $\alpha _n(h)$ are eigenvalues of the nonlinear spectral problem (16). 
As an example, the dependences of $V_n(h)$ for the soliton complex and its 
"excited" states at $\beta =1/12$ are pictured in Fig. 14. Every branch has 
a finite range of the velocity changing from the maximum values 
$V_n^0=\sqrt{1-(\beta /\alpha _n^0)^{1/2}}$ to zero. It is evident
that critical fields $h_n$ corresponding to static solutions are found from 
the equation $\alpha _n(h_n)=\beta$. The number of possible complex states 
is finite because the states with $\alpha _n(h_n)<\beta$ turn out to be 
forbidden. 

The regularized 2dDSG equation (10), at any $\beta$, has a complete 
infinite series of soliton complex states with the following allowed
velocities: 

\begin{equation}
\label{vnh}V_n(h)=\sqrt{1+\frac \beta {4\alpha _n(h)}}-%
\sqrt{\frac \beta{4\alpha _n(h)}} .
\end{equation}
In the case of $n=1$ we have the explicit expression for 
$\alpha _1(h)$ (Eq. (74)), and after its substitution to Eq. (83) we
come back to Eq. (20). It is clear that the maximum velocity value is 
realized at $h=0$, when $\alpha=3/4$; it coincides with $V_r(\beta)$ of
Eq. (7). At high $h$ and hence small $\alpha _n(h)$ the velocity 
dependences are simplified to $V_n(h)\simeq\sqrt {{\alpha _n(h)}/\beta}$.

In conclusion we present the energies of the complex and two "excited"
states as functions of $h$ for a prescribed $\beta =1/12$ (Fig. 15). As 
shown above (see Eq. (21)), $E_1$ is independent of $h$. Energies of the 
"excited" states turn out to be less than $E_1$; therefore this terminology 
becomes inadequate when $\beta$ is fixed. It seems also that the 
energy arguments about the complex stability do not work in this 
case. However, it should be noted that the velocities of all the 
complex states are different, and the exact complex solution possesses 
the highest velocity. A stability criterion has to be formulated so 
that the energy comparison is to be performed under the condition of 
the conservation of the another integral of motion, the momentum. Since 
for a fixed $\beta$, both first integrals are functions of the  
parameter $h$ alone, then for a given $h$, the momenta are different, 
so that such a simple comparison becomes impossible. The understanding of 
the above studied mechanisms of the formation of the topological soliton 
complexes allows us to believe in their stability in the dispersive 
conservative systems.

\bigskip

\begin{center}
{\bf 7. SUMMARY AND DISCUSSION}
\end{center}

\bigskip

In this section we discuss some peculiarities of soliton-complex dynamics
in strongly dispersive models and formulate final general conclusions.

Dispersive nonlinear equations with fourth-order spatial derivatives, 
examples of which have been studied above, are used usually as a first 
approximation in the description of discrete systems. The present 
investigation shows that the higher dispersive terms in the nonlinear 
wave equations effect much more essentially than the small perturbations.
The dispersion causes the strong dissipation of energy of the moving 
$2\pi$-soliton, but also makes it possible the creation of the bound soliton
complexes consisting of two or more $2\pi$-solitons which can move 
radiationlessly. 

The formation of soliton complexes turns out to be the universal 
phenomenon in nonlinear strongly dispersive media, which are of 
both theoretical and practical interest.

From the theoretical point of view it is very interesting that the dispersion
produces a discrete spectrum for a nonlinear eigenvalue problem for purely
solitonic solutions, and its influence is not reduced simply to small changes
of the $2\pi$-kink shape as it was believed before. Analytical and numerical
considerations confirm the absence of the steady $2\pi$-kink solution in the
above-studied dispersive systems with the fourth and higher order spatial
derivatives. The earlier papers (see the review $[49]$ for references)
declared the existence of such solutions. Indeed, application of the
perturbation theory, using the smallness of $\alpha$, leads to this erroneous
conclusion. It seems to be possible to construct formal asymptotic series for
the $2\pi$-kink solutions of Eqs. (16) and (72). However, summation of all
terms of the series results in disappearance of the periodicity of the
function $F(u)$ with period $2\pi$, and only solutions with periods $4\pi,
6\pi ...$ survive.

At the same time these peculiarities of the dispersive effects require 
some caution when one exploits the variational description of the 
bound solitons or breather states in dispersive or discrete 
systems $[50]$.

The numerical finding of the discrete set of periodic eigenfunctions in the 
Eq. (72) could point to the possibility of a complete integrability 
of the equation, at least, in some special cases. However this question 
is still open.

The existence of soliton complexes consisting of three or more solitons is not
discussed in detail in the given paper. The above analysis of the mechanisms
and conditions of the creation of the complexes is applicable to the case of
the multisoliton bunching. The presence of exact solutions in Eqs. (24) and
(27) may be used as a basis for further studies in this direction. One is
easily convinced that the exact solutions found are not exotic. Specific
relations between constants $\beta$ and $\gamma$ of Eqs. (23) and (26) are
required only as conditions of the existence of the solution in a very simple
analytical form. If we slightly alter the value of the parameter $\gamma$, the
solution still exists, just as before.  Numerical integration of the dSG
equation in the stationary case $[11]$ revealed such a solution for 
$\gamma=0$. Hence the multicomplex solutions are stable with respect to 
changing the dispersive parameters.

The important question of the stability of the soliton complexes in the 
framework of the theory of partial differential equations has remained beyond 
the scope of the paper. Work in this direction is in progress. However, 
previous studies of the problem in the discrete and nonlocal 
sine-Gordon models suggest a positive answer to this question.

From the experimental point of view the soliton complexes may be very 
attractive for the use in energy and information transfer processes. 
In particular, the discrete arrays of Josephson junctions can be  
described by one-dimensional dispersive sine-Gordon models $[30,31,50]$.
The fluxons, the $2\pi$-kinks, in these systems could form the 
soliton complexes with specific properties discussed above.
Therefore, Josephson junction arrays may be considered as suitable 
candidates for real experiments, in which the soliton complexes would 
manifest themselves $[15]$.
Other examples of appropriate physical systems are the low-dimensional  
ferro- and antiferromagnets. More definitely, it may be an one-dimensional
biaxial ferromagnet with strong easy-plane anisotropy and a magnetic 
field applied along the plane. This system is described by the double 
sine-Gordon equation, and taking into account the discreteness effects 
duly leads to the dispersive model (9). Two-dimensional antiferromagnets 
with the weak interplane exchange can be also treated in the framework of 
the models analyzed in previous sections. The nonlinear dynamics of 
dislocations is another potential field of application of the obtained 
results $[1,2]$. Thus there are many experimental possibilities for the 
observation of the occurrence of stationary soliton-complex dynamics 
in dispersive media.

In conclusion we note that the present investigation shows that purely 
soliton complexes can be realized in dispersive media which are 
modulationally unstable from the very beginning, as well as in systems 
where the $2\pi$-kink motion produces the medium modulation due to the 
radiation accompanying the soliton movement. In both situations 
nonstationary radiative dynamics of two solitons can result in the final 
formation of the purely solitonic complex with cancelling the oscillations
on soliton wakes, what can be considered as a specific interference 
effect. However, discussion of the nonstationary soliton complex 
dynamics as well as the problem of their dynamics in dissipative systems 
is the subject matter of following publications.

\bigskip

\bigskip

\begin{center}
{\bf ACKNOWLEDGEMENTS}
\end{center}

The authors are grateful to A.S.Kovalev and I.M.Babich for fruitful 
discussions. This research was supported by INTAS\ grant (No.93-1662). 
M.B. thanks for hospitality Laboratoire de Mod\'{e}lisation en 
M\'{e}canique (UMR CNRS 7607), Universit\'{e} Pierre et Marie Curie, 
Paris, France and acknowledges for support from the Visitors Program 
of Max-Planck-Institut f\"{u}r Physik Komplexer Systeme, 
Dresden, Germany, where parts of this reseach were carried out.

\bigskip

\bigskip

\begin{center}
{\bf APPENDIX }
\end{center}

\begin{center}
{\bf A. Effective Lagrangian and Hamiltonian}
\end{center}
After inserting the anzatz (31) into the Lagrangian (11) 
and the Hamiltonian (12) we perform  all needed integrations 
using the following values of intergals:

$$
\begin{array}{rcccl} 
I_{0}(R) & = & \int\limits_{-\infty }^\infty dy\frac{\cosh^2(R)}{\cosh^2(R)+%
\sinh^2(y)} & = & 2R\coth(R)\\
I_{1}(R) & = & \int\limits_{ -\infty }^\infty dy (\frac{\cosh(R)\cosh(y)}%
{(\cosh^2(R)+\sinh^2(y)})^2 & = & 1+\frac { \D 2R}{\D \sinh(2R)}\\
I_{2}(R) & = & \int\limits_{ -\infty }^\infty dy (\frac{\cosh(R)\sinh(y)}%
{\cosh^2(R)+\sinh^2(y)})^2 & = & {\coth^2(R)}(1-%
\frac { \D 2R}{\D \sinh(2R)})\\
I_{3}(R) & = & \int\limits_{ -\infty}^\infty dy (\frac{\cosh(R)y}{\cosh^2(R)%
+\sinh^2(y)})^2 & = & {\frac{\D 1}{\D 3}}(R^2+%
{\frac{\D \pi^2}{\D 4}})I_{0}(R)\\
I_{4}(R) & = & \int\limits_{ -\infty}^\infty dy (\frac{y\sinh(y)\cosh(R)}%
{\cosh^2(R)+\sinh^2(y)})^2 & = & {\frac{\D 1}{\D 3}}[2R^2\coth^2(R)+%
(R^2+{\frac{\D \pi^2}{\D 4}})I_{2}(R)] .
\end{array}
$$
As a result expressions for the kinetic and potential energies of the 
two-soliton system can be written as

\begin{equation}
\label{kin}
\begin{array}{c}
T={\frac {\D 8}{\D l}}\{{\frac {\D 1}{\D 3}}{l_t}^2[2{R^2}+(R^2+%
{\frac{\D\pi^2}{\D 4}})I_{1}(R)]+%
(l{R_t}\tanh(R))^2I_{2}(R)+2R{R_t}ll_t\\
+{X_t}^{2}I_{1}(R)\} ,
\end{array}
\end{equation}
and

\begin{equation}
\label{pot01}
\begin{array}{c}
U={\frac {\D 8}{\D l}}\{I_{1}(R)+l^2I_{2}(R)+ 2hl^{2}I_{0}(R)\\
-{\frac{\D \beta}{\D l^2}}[{\frac {\D 1}{\D 3}}-{\tanh^2(R)}(1-%
I_{2}(R)(1+{\frac{\D 2}{\D \sinh^2(2R)}}))]\} .
\end{array}
\end{equation}
Then the effective Lagrangian and Hamiltonian are expressed as usual

\begin{equation} 
\label{lagha}L_{eff}=T-U,\hspace{1cm}H_{eff}=T+U .
\end{equation}
The Lagrangian function with accuracy to order $O(R^2)$ can be written

\begin{equation}
\label{FullL}L=L_0+R^2[T_1-U_1(q,X_{t}^2)] ,
\end{equation}
where the zero-order Lagrangian is the following one:

\begin{equation}
\label{Lo}L_0=\frac{\D 16}{\D l} \{\frac{\pi ^2}{12}l_t^2 -1+X_{t}^2-%
(\frac 1 3+h)l^2+\frac \beta{3l^2}\} ,
\end{equation}
and $T_1$ contains all terms proportional to the time derivatives of
function $l$ and  $U_1$ is given by:

\begin{equation}
\label{Up}U_1(q,X_{t}^2)=\frac {16}{3l} \{X_{t}^2 -1+\frac {1^2} 5 +%
\frac 7 5 \frac \beta l^2\} .
\end{equation}
From Eq. (87) we see that one of the Lagrange equation takes the
form $\partial L/\partial R=0$. 
It is transformed into $RU_1(q,X_{t}^2)=0$ for stationary states. 

\bigskip

\begin{center}
{\bf B. Intergable sine-Gordon limit. Exact two-soliton solution}
\end{center}

If we put $h=0$, $\beta=0$, $l=\gamma^{-1}(v)\equiv\sqrt{1-v^2}=const$, and
$X_t=0$, then we reduce the problem to the consideration of the sine-Gordon 
system in the reference frame moving with mass center. In this case the 
Hamiltonian $H_{eff}$ is simplified to $H_{SG}$:

\begin{equation}
\label{excsg}H_{SG}=\frac 8{\sqrt{1-V^2}} \{ (1+%
(\tanh(R) R_t)^2)I_{2}(R)\gamma^{-2}+I_{1}(R) \} .   
\end{equation}
It is easy to be convinced that the function $R(t)$

\begin{equation} 
\label{lagham01}R(t)=\ln[{\frac1v}{\cosh({\gamma}vt)}+%
\sqrt{({\frac1v}{\cosh({\gamma}vt)})^2-1}]
\end{equation}
is the exact solution of the effective Hamiltonian equations. The
substitution of $R(t)$ into the ansatz (31) and $H_{SG}$ provides the 
exact two-soliton solution of the integrable SG equation and 
corresponding value for the energy $E_0=16\gamma$. 

\bigskip

\begin{center}
{\bf C. Double sine-Gordon limit. Small wobbler oscillations}
\end{center}
Suppose again that $\beta=0$ but let the parameter $l$ be the function 
of $R$ of the form  $l(R)=\tanh(R)$. Then after substitution of 
$l(R)$ into the Lagrangian we find for the double sine-Gordon system:

\begin{equation}
\label{lag5}
L_{DSE}=8{R_t}^2\tanh(R)G(R)-\coth(R)-hR .
\end{equation}
The last two terms in Eq. (92) correspond to the potential energy, and 
the function $G(R)$ is introduced as

\begin{equation}
\label{lagl}
G(R)=I_{1}(R)+{\frac4{\D 3\sinh^2(2R)}}[2{R^2+%
I_{1}(R)(R^2+{\frac{\D \pi^2}{\D 4}})]} .
\end{equation}
The stationary solution is found from the condition of minimum of potential
energy. One determines the equilibrium value of the parameter $R$ from the
equation $\sinh(R_w)=1/\sqrt{h}$. Naturally, the ansatz coincides with the
exact wobbler solution (15). 

Small oscillations near the equlibrium position $r(t)=R(t)-R_w$ are 
described by a linear equation which is found from the Lagrangian by 
the usual way. 

\begin{equation}
\label{lagl01}
r_{tt}+\Omega^2r=0 .
\end{equation}
The frequency is given as
\begin{equation}
\label{ome}\Omega^2(h)=2h\coth^2(R_w(h))/G(R_w(h)) .
\end{equation}

This field dependence for the frequency of the internal oscillations 
is seemed to be the best analytical approximation of this function 
for the moment (see $[51]$ and references therein for the comparison). 

\bigskip

\bigskip

\begin{center}
{\bf REFERENCES}
\end{center}

\small

\noindent 1. N. Flytzanis, S. Crowley, and V. Celli, Phys.Rev.Lett. 
{\bf 39}, 891 (1977).

\noindent 2. M. Peyrard and M.D. Kruskal, Physica D,{\bf 14}, 88 (1984).

\noindent 3. M. Peyrard, S. Pnevmatikos, and N. Flytzanis, Physica D {\bf
19}, 268 (1986).

\noindent 4. R. Boesch, C.R. Willis, and M. El-Batanouny, Phys.Rev.B
{\bf 40}, 2284 (1989).

\noindent 5. M.M. Bogdan, A.M. Kosevich, and V.P. Voronov,  
In {\em Soliton and Applications}, edited by V.G.Makhankov, V.K.Fedyanin,
and O.K.Pashaev (World Scientific, Singapore, 1989), pp.231-243.

\noindent 6. P.T. Dinda and C.R. Willis Phys.Rev.E {\bf 51}, 4958 (1995).

\noindent  7. S. Flach and C.R. Willis, Phys. Rep. {\bf 295}, 181 (1998).

\noindent 8. K. Nakajima, Y. Onodera, T. Nakamura, and R. Sato, 
J. Appl. Phys, {\bf 45,} 4095 (1974).

\noindent 9. K.A. Gorshkov, L.A. Ostrovsky, and V.V. Papko, Sov.Phys. 
JETP {\bf 44}, 306 (1976).

\noindent 10. K. Lonngren,  In {\em Solitons in Action }, edited by
K.Lonngren and A.C.Scott (Academic Press, New York, 1978), 
pp.138-162.

\noindent 11. G.L. Alfimov, V.M. Eleonskii, N.E. Kulagin, and N.N. 
Mitskevich, Chaos, {\bf 3}, 405 (1993).

\noindent 12. L.Vazquez, W.A.B.Evans, and G.Rickayzen, Phys.Lett. A  
{\bf 189}, 454 (1994).

\noindent 13. Yu. Gaididei, N.Flytzanis, A. Neuper, and F.G.Mertens, 
Phys. Rev. Lett., {\bf 75}, 2240 (1995).

\noindent 14. G.L. Alfimov and V.G.Korolev, Phys.Lett. A {\bf 246}, 429 
(1998).

\noindent 15. A.V. Ustinov, B.A. Malomed, and S. Sakai, Phys.Rev. B 
{\bf 57}, 11691 (1998)

\noindent 16. T. Kawahara, J.Phys.Soc.Japan, {\bf 33}, 260 (1972).

\noindent 17. N.Yajima, M.Oikawa, and J.Satsuma, Journ. Phys.
Soc. Jap., {\bf 44}, 1711 (1978).

\noindent 18. T.Kawahara and S.Toh, Phys. Fluids, {\bf 31}, 2103 (1988).

\noindent 19. N.A. Zharova and A.M. Sergeev, Sov.J.Plasma Phys. 
{\bf 15}, 681 (1989).

\noindent 20. A. H\"{o}\"{o}k and M.Karlsson, Opt.Lett. {\bf 18}, 
1390 (1993). 

\noindent 21. V.I.Karpman, Phys.Lett. A, {\bf 193}, 355 (1994).

\noindent 22. C.I.Christov, G.A.Maugin, and M.G.Velarde. 
Phys. Rev. E, {\bf 54}, 3621 (1996). 

\noindent 23. V.E.Zakharov and E.A.Kuznetzov, ZhETF, {\bf 113}, 
1892 (1998).

\noindent 24. K.A.Gorshkov and L.A.Ostrovsky, Physica D {\bf 3}, 428 
(1981).

\noindent 25. B.A. Malomed, Phys.Rev. E {\bf 47}, 2874 (1993).

\noindent 26. A.V.Buryak and N. N. Akhmediev, Phys.Rev. E {\bf 51}, 
3572 (1995).

\noindent 27. A.V.Buryak, Phys. Rev. E, {\bf 52} 1156 (1995).

\noindent 28. D.W.McLaughlin and A.C.Scott, Phys.Rev.A, {\bf 18},
1652 (1978).

\noindent 29. V.I.Karpman and  V.V.Soloviev, Physica D, {\bf \ 3},
142 (1981).

\noindent 30. M.M.Bogdan and A.M.Kosevich,   In
{\em Nonlinear Coherent Structures in Physics and Biology },
edited by K.H.Spatschek and F.G.Mertens, NATO ASI Series, 
Physics {\bf 329} (Plenum Press, New York, 1994), pp.373-376.

\noindent 31. M.M.Bogdan and A.M.Kosevich , Proc. Estonian Acad. Sci. 
Phys. Math., {\bf 46}, 14 (1997).

\noindent 32. G.A.Maugin and C.I.Christov, Proc. Estonian Acad. Sci. 
Phys. Math., {\bf 46}, 78 (1997).

\noindent 33. G.A.Maugin, Proc. Estonian Acad. Sci. Phys. Math.,
{\bf 44}, 40 (1995).

\noindent 34. A.M.Kosevich and A.S.Kovalev, Solid. State Commun. 
{\bf 12}, 763 (1973).

\noindent 35. A.V.Savin, A.V. Sov. JETP, {\bf 108}, 1105 (1995).

\noindent  36. O.V.Hendelman, L.I.Manevich, ZhETF, {\bf 112}, 1510 (1997).

\noindent 37. Yu.S. Kivshar, A.R.Champneys, D.Cai, and A.R.Bishop,
Phys.Rev.B, {\bf 58}, 5423 (1998).

\noindent  38. A.V.Ustinov, M.Cirillo, and B.A.Malomed, Phys.Rev.B {\bf 47}, 
8357 (1993).

\noindent 39. M. Peyrard, B.Piette, and W.J.Zakrzewski, Physica D {\bf 64},
355 (1993).

\noindent 40. G.L.Alfimov, V.M.Eleonskii, and N.V.Mitskevich, JETP {\bf 76},
563 (1993).

\noindent 41. K.M.Leung, Phys.Rev.B, {\bf 26}, 226 (1982). 

\noindent 42. C.A.Condat, R.A.Guyer, and M.D.Miller, Phys.Rev.B, 
{\bf 27}, 474 (1983). 

\noindent 43. G.A.Maugin and A.Miled, Phys.Rev.B, {\bf 33}, 4830 (1986).

\noindent 44. P.Rosenau, Phys.Rev.B, {\bf 36}, 5868 (1987).

\noindent 45. M.M.Bogdan, A.M.Kosevich, and G.A.Maugin, Cond.Matt.Phys., 
{\bf 2}, N1 (17), 1 (1999).

\noindent 46. M.Karlsson and A. H\"{o}\"{o}k , Opt. Commun. {\bf 104}, 
303 (1994). 

\noindent 47. L.D.Landau and E.M.Lifshitz. Quantum Mechanics (Pergamon,
New York, 1977).

\noindent 48. M.M.Bogdan (to be published).

\noindent 49. Yu.Kivshar and B.A.Malomed, Rev. Mod. Phys., {\bf 61}, 
763 (1989).

\noindent 50. J.A.D.Wattis, Nonlinearity, {\bf 9}, 1583 (1996).

\noindent 51. C.R.Willis, M.El-Batanouny, and R.Boesch,
Phys.Rev.B, {\bf 40}, 686 (1989).

\bigskip

\bigskip

\normalsize

\begin{center}
{\bf FIGURES CAPTIONS}
\end{center}

Fig. 1. Velocities of soliton complexes as functions of the 
discreteness parameter $d$. Solid (dash) line corresponds to the 1dSG 
(2dSG) equation.

\medskip

Fig. 2. The potential shapes of the dispersive double sine-Gordon and 
piecewise-linear models.

\medskip

Fig. 3 (a)-(d). The potential energy of the dispersive double 
sine-Gordon system in the variational approach as the function of the 
effective length and the separation between solitons. Figs. (a) and 
(b) correspond to $\beta=1/5$, and (c) and (d) to $\beta=3/4$, 
respectively. The parameter $h=0$ for the cases (a) and (c), and 
$h=0.1$ for the cases (b) and (d).

\medskip

Fig. 4. Eigenvalues $\alpha _n$ ($n=1..10$) corresponding to the 
soliton complexes. Solid circles denote results of the 
phenomenological approach to the stationary dispersive sine-Gordon 
equation. Open circles correspond to the piecewise-linear model. Solid 
squares are the eigenvalues obtained by the numerical integration of 
the dSG equation.

\medskip

Fig. 5. Analytical dependences $\alpha _n(h)$ ($n=1..5$) found after 
one step of the phenomenological iteration procedure for the dDSG 
equation.

\medskip

Fig. 6. Dependences $\alpha _n(h)$ ($n=1..5$) obtained for the 
dispersive piecewise-linear model.

\medskip

Fig. 7. Five first soliton complex eigenfunctions constructed 
explicitly in the piecewise-linear model (the case $h=0$). 

\medskip

Fig. 8. Numerical results for the eigenvalues $\alpha _n(h)$ 
($n=1..5$) in the dDSG equation.

\medskip

Fig. 9. The soliton complex and its first four "excited" states found 
numerically for the dSG stationary equation.

\medskip

Fig. 10 (a),(b). "Phase portraits" of soliton complex and the 
odd (a) and even (b) "excited" states in the dSG equation.

\medskip

Fig. 11. The soliton complex and its three "excited" states found 
numerically for the dDSG equation ($h=0.1$).

\medskip

Fig. 12 (a),(b). "Phase portraits" of soliton complex and the odd (a) 
and even (b) "excited" states in the dDSG equation ($h=0.1$).

\medskip

Fig. 13. Energy dependences of the soliton complex and the "excited" 
states on the parameter $h$. The energies are normalized by the 
wobbler energy $E_w(h)$.

\medskip

Fig. 14. Velocities of the soliton complex and the "excited" states 
as functions of the parameter $h$. The dispersive parameter $\beta$
equals $1/12$.

\medskip

Fig. 15. Energies of the soliton complex and two "excited" states 
as functions of the parameter $h$. The lines finish at critical 
values $h_n$. The value $h_1=4/3$ is beyound the figure domain.
The parameter $\beta$ is fixed as $1/12$.

\end{document}